\documentclass[twocolumn,showpacs,preprintnumbers,amsmath,amssymb,superscriptaddress]{revtex4-1}
\usepackage{graphicx}

\begin{document}
\title{Multiscale modeling of oscillations and spiral waves in \emph{Dictyostelium} populations}

\author{ Javad Noorbakhsh}
\affiliation{Physics Department, Boston University, Boston, MA, USA}

\author{David Schwab}
\affiliation{Physics Department, Princeton University, Princeton, NJ, USA}

\author{Allyson Sgro}
\affiliation{Physics Department, Princeton University, Princeton, NJ, USA}

\author{Thomas Gregor}
\affiliation{Physics Department, Princeton University, Princeton, NJ, USA}

\author{Pankaj Mehta}
\email{pankajm@bu.edu}
\affiliation{Physics Department, Boston University, Boston, MA, USA}

\date{\today}
\begin{abstract}
Unicellular organisms exhibit elaborate collective behaviors in response to environmental cues. These behaviors are controlled by complex biochemical networks within individual cells and coordinated through cell-to-cell communication.  Describing these behaviors requires new mathematical models that can bridge scales -- from biochemical networks within individual cells to spatially structured cellular populations. Here, we present a family of multiscale models for the emergence of spiral waves in the social amoeba {\it Dictyostelium discoideum}. Our models exploit new experimental advances that allow for the direct measurement and manipulation of the small signaling molecule cAMP used by {\it Dictyostelium} cells to coordinate behavior in cellular populations. Inspired by recent experiments, we model the {\it Dictyostelium} signaling network as an excitable system coupled to various pre-processing modules. We use this family of models to study spatially unstructured populations by constructing phase diagrams that relate the properties of  population-level oscillations to parameters in the underlying biochemical network. We then extend our models to include spatial structure and show how they naturally give rise to spiral waves. Our models exhibit a wide range of novel phenomena including a density dependent frequency change, bistability, and dynamic death due to slow cAMP dynamics.  Our modeling approach provides a powerful tool for bridging scales in modeling of {\it Dictyostelium} populations.
\end{abstract}

\maketitle

\section{Introduction}

Collective behaviors are ubiquitous in nature. They can be observed in diverse systems such as animal flocking, microbial colony formation, traffic jamming, synchronization of genetically engineered bacteria and social segregation in human populations \cite{Ballerini2008,Ben-Jacob2000,Flynn2009,Danino2010,Schelling1971}. A striking aspect of many of these systems is that they span a hierarchy of scales and complexity.  A common property of such complex systems is that the collective behavior at larger scales can often be understood without a knowledge of many  details at smaller scales. This important feature allows one to study the system on multiple distinct spatiotemporal scales and use the information obtained in each scale to develop a more coarse-grained model at a larger scale. This approach has been termed multi-scale modeling and provides a framework for the study of complex systems \cite{Meier-Schellersheim,Qutub2009,Noble2008}. Compared to a fully detailed modeling approach, multi-scale models are more amenable to computer simulations, contain fewer ad hoc parameters and are easier to interpret. As a result these models can be very useful in developing theoretical understanding of complex systems. For example, great success has been achieved in study of pattern formation in microbial colonies by modeling them as a continuum of cells with simple rules such as growth and diffusion \cite{Ben-Jacob2000}. 

One interesting system that exhibits different behaviors on different spatiotemporal scales is the social amoeba {\it Dictyostelium discoideum} \cite{Mehta2010}.  {\it Dictyostelium} has a fascinating lifecycle. It starts as a population of unicellular organisms that can separately grow and divide. However, when starved, these cells enter a developmental program where individual cells aggregate, form multicellular structures, and eventually, fruiting bodies and spores. Upon starvation, cells  produce the small signaling molecule cAMP, and excrete it into the environment in periodic bursts. Each cell responds to an increase in the concentration of extracellular cAMP by secreting more cAMP, resulting in pulses that propagate through the environment as spiral waves.  Cells eventually aggregate at the center of these spiral waves and  search for food collectively \cite{McMains2008}. 
In addition to its fascinating life cycle,  {\it Dictyostelium} is also an important model organism for eukaryotic chemotaxis. The {\it Dictyostelium} chemotaxis network  is highly conserved among eukaryotes \cite{Swaney2010}, and is thought to be a good model for many medically relevant phenomena ranging from neutrophil chemotaxis and cancer metastasis to cell migration during animal morphogenesis \cite{Parent2004, Zernicka-Goetz2006}.

There has been extensive work on modeling the {\it Dictyostelium} signaling network, starting with the pioneering work by Martiel {\it et al.} \cite{Martiel1987}. These authors suggested that oscillations and spiral waves emerge from negative feedback based on desensitization and adaptation of the cAMP receptor. More recent models extend on this work by incorporating additional proteins  known to play a significant role in the {\it Dictyostelium} signaling network \cite{Laub1998}. Although very successful at producing oscillations and spiral patterns, these models are inconsistent with recent quantitative experiments that show cells oscillate even in the presence of saturating levels of extracellular cAMP \cite{Gregor2010}. Other models have focused on reproducing the eukaryotic chemotaxis network, which shares many molecular components with the signaling network responsible for collective behavior \cite{Takeda2012,Wang2012}. These models explore how cells respond to an externally applied pulse of cAMP but do not attempt to model oscillations or spiral waves. Combinations of such models with oscillatory networks represent a possible route for multiscale modeling \cite{Xiong2010} but have not been extensively studied.  Other models have focused on reproducing spiral waves in {\it Dictyostelium} populations using reaction diffusion equations and cellular automata \cite{Aranson1996,Kessler1993}.  While these models tend to be very successful at producing population level behaviors, it is hard to relate these models to the behavior of single cells. This highlights the need for new mathematical models that can bridge length and complexity scales.

Recently, there have been tremendous experimental advances in the study of {\it Dictyostelium}. Using microfluidics and single-cell microscopy,  it is now possible to produce high-resolution time-course data of how single {\it Dictyostelium} cells respond to complex temporal cAMP inputs \cite{Gregor2010,Xiong2010,Wang2012,Cai2011,Song2006,Sawai2007,Cai2010,Masaki2013}. By combining such quantitative data with ideas from dynamical systems theory and the theory of stochastic processes,  we recently \cite{Sgro2014} proposed a new universal model for the {\it Dictyostelium} signaling network, based on an excitable signaling network coupled  to a logarithmic ``pre-processing" signaling module (see Figure \ref{fig:Schematic}). To make a phenomenological model for single and multicellular behavior we exploited the observation that the {\it Dictyostelium} signaling network is poised near a bifurcation to oscillation. Each {\it Dictyostelium} cell was treated as an excitable FitzHugh-Nagumo model that was coupled to other cells through the concentration of the extracellular cAMP.  A central finding of this model was that intracellular noise is a driving force for multicellular synchronization. 

Inspired by these results, in this paper we analyze a family of models for cells communicating via an external signal such as cAMP. The external signal is detected by the cell, transduced through a preprocessing module which can  be linear, logarithmic, or Michaelis-Menten,  and then fed into an excitable signaling network. Using these models, we explore the rich population-level behaviors that emerge in coupled oscillator systems  from the interplay of stochasticity, excitability, and the dynamics of the external medium. We also extend our models to include space and show that spiral waves naturally emerge in the limit of large population densities. In contrast to earlier models for spiral waves, we can explicitly include the dynamics of  extracellular cAMP and treat it distinctly from the dynamics of signaling networks. 

Our model naturally overlaps with, and complements, the extensive literature of coupled oscillatory and excitable systems. Coupled oscillators have been observed in many different biological systems such as neuronal networks, circadian rhythm, Min system and synthetic biological oscillators \cite{Traub1989,Enright1980,Mirollo1990,Meinhardt2001,Danino2010}. Most theoretical models focus on directly coupled oscillators and relatively little work has been done on noisy oscillators coupled through a dynamical external medium such as cAMP  \cite{Schwab2012a,Schwab2012}. Furthermore, an important aspect of our model is the role played by stochasticity. It is well-known that noisy systems are not easily amenable to traditional methods in dynamical systems theory \cite{Tanabe2001,Lindner2004} and concepts such as bifurcation point are ill-defined in this context. For this reason, the {\it Dictyostelium} signaling network provides a rich, experimentally tractable system for exploring the physics of noisy oscillators coupled through an external medium.

The paper is organized as follows. We start by introducing our family of models. We then construct  phase diagrams  describing the behavior of spatially-homogenous populations, focusing on the regime where extracellular signaling molecules are degraded quickly compared to the period of oscillations. We then analyze the opposite regime where signaling dynamics is slow and show that this gives rise to novel new behaviors such as dynamic death. Finally, we extend the model to spatially inhomogeneous populations and study how spiral waves naturally arise in these models. We then discuss the biological implications of our results, as well as,  the implications of our model for furthering our understanding of coupled oscillators.

\section{ Modeling \emph{Dictyostelium} Populations}
\label{sec:Model}

\begin{figure}
\centering
\includegraphics[width=0.8\linewidth]{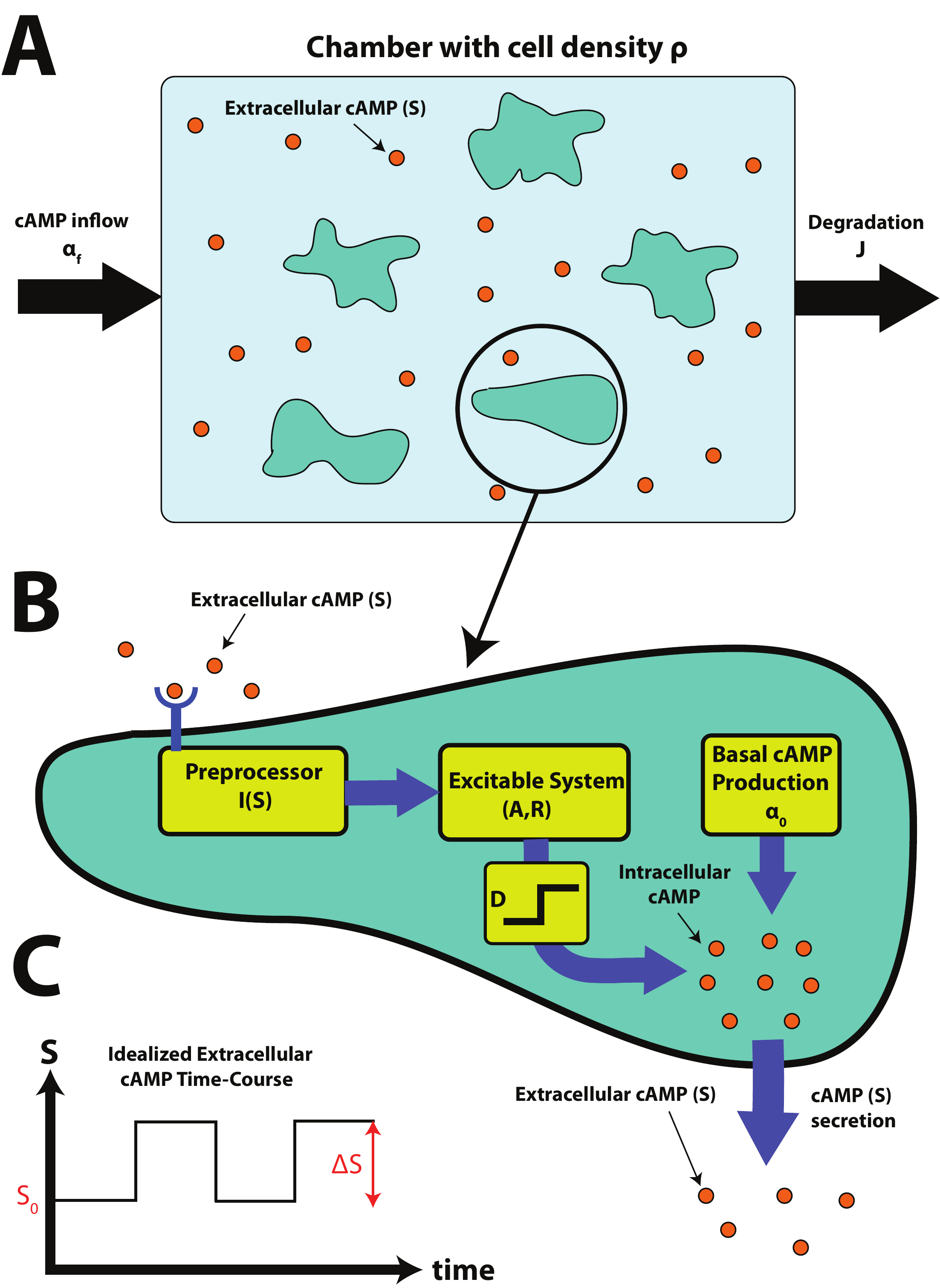}
\caption{\textbf{Model Schematic - } {\bf A)} Schematic of the experimental setup. A population of {\it Dictyostelium discoideum} cells with density $\rho$ is placed in a microfluidic chamber. cAMP is flown into the chamber with rate $\alpha_f$ and the medium is washed out with rate $J$. The concentration of extracellular cAMP is labeled by $S$. \textbf{B)} Schematic of cell model. Extracellular cAMP concentration ($S$) is detected by the cell and preprocessed through the function $I(S)$. The result is fed into an excitable system with internal variables $A$ and $R$. The value of $A$ is then thresholded and amplified by $D$ to produce more cAMP for secretion. Simultaneously cAMP is also being produced with a constant rate $\alpha_0$ and leaks into the extracellular environment. {\bf C)} An idealized time-course of extracellular cAMP concentration ($S$) is shown  in the large $J$ regime where the concentration changes according to a square wave with baseline $S_0$ and amplitude  $\Delta S$. We refer to $S_0$ and $\Delta S$ as the background cAMP and firing-induced cAMP, respectively.}
\label{fig:Schematic}
\end{figure}

New experimental advances allow for the direct measurement and manipulation of the small signaling molecule cAMP used by {\it Dictyostelium} cells to coordinate behavior in cellular populations. In such experimental systems, a few hundred {\it Dictyostelium} cells are confined in a microfluidic device. The levels of intracellular cAMP within cells can be measured quantitatively using a F\"{o}rster Resonance Energy Transfer (FRET)-based sensor \cite{Gregor2010,Sgro2014}. This allows for precise, quantitative measurements of the response of the {\it Dictyostelium} signaling networks to complex temporal signals of extracellular cAMP.  Cells are placed in a microfluidic device at a density $\rho$. The microfluidic device allows for rapid mixing and exchange of extracellular buffer, which ensures that cells experience a uniform and controlled environment.  The flow rate of buffer can be experimentally manipulated. Large flows wash away the extracellular cAMP produced by cells, resulting in a larger effective degradation rate, $J$, for extracellular cAMP. It is also possible to add cAMP to the buffer at some rate $\alpha_f$. This experimental set-up is summarized in Figure \ref{fig:Schematic}. 
 
 We start by building models for spatially unstructured populations where the extracellular cAMP concentration is assumed to be uniform. In this case, all cells in the chamber sense the same extracellular cAMP concentrations and we can ignore all spatial effects. To model individual cells, we build upon our recent work  \cite{Sgro2014} where we showed that the dynamics of the {\it Dictyostelium} signaling network  can be modeled using a simple, universal, excitable circuit: the noisy Fitzhugh-Nagumo (FHN) model. To realistically model the {\it Dictyostelium} signaling circuit, it is necessary to augment the FHN with an additional ``pre-processing'' module that models the signal transduction of extra-cellular cAMP levels upstream of this core excitable circuit (Figure \ref{fig:Schematic}B).  In the full signaling circuit, extracellular cAMP is detected by receptors on cell membrane. The resulting signal is funneled through several signal transduction modules, ultimately resulting in the production of cAMP. To model this complicated signal transduction process, we use a family of preprocessing modules, whose output serves as an input into the universal excitable circuit. 
 
 Inspired by the {\it Dictyostelium} circuit, we assume that the dynamics of the preprocessing module are fast compared to the excitable dynamics of cAMP signaling circuit. For example, the typical time scale associated with the early signaling protein Ras is of order 30 seconds whereas cAMP oscillations have periods of order 300 seconds \cite{Takeda2012,Gregor2010}. This allows us to model the preprocessing modules using a monotonically increasing function, $I(S)$, that relates the output of the preprocessing module to the extracellular cAMP concentration, $S$. In this work, we will consider three different biologically inspired pre-processing modules:  (1) a linear module $I(S)=S$ where the extracellular cAMP signal does not undergo any  preprocessing; (2) a Michaelis-Menten module, 
 \begin{equation}
 I(S)=\frac{\beta S}{S+K_D},
 \end{equation}
 where the output is a saturating function of the extracellular cAMP;  and (3) the logarithmic module that senses fold changes
 \begin{equation}
 I(S)=a\log{\left(1+S/K\right)}.
 \end{equation}

The output of these modules is fed into a universal, excitable circuit modeled by the FHN. The FHN model consists a set of inter-locking positive and negative feedback loops consisting of an activator, $A$, that quickly activates itself through positive feedback, and on a slower time scale, activates a repressor $R$, that degrades the activator $A$. The FHN model is the prototypical example of an excitable system, and can spike or oscillate depending on the external input. To incorporate the biology of cAMP secretion by {\it Dictyostelium} cells  in response to external inputs, we assume that when a cell spikes, it releases cAMP into the environment. To determine when a cell spikes, we threshold the activator variable $A$  using a Heaviside function $\Theta(A)$, where $\Theta(x)=1$ if $x>0$ and $\Theta(x)=0$ if $x=0$. Finally, we assume that cells produce and secrete cAMP at a spike-independent basal rate, $\alpha_0$. This can be summarized by the equations
\begin{align}
\label{eqn:Model}
\frac {dA_i}{ dt} &= A_i-\frac{1}{ 3}{A_i}^3-R_i+I(S)+\eta_i(t) , \qquad i = \{1,2,...,N\}\\\nonumber  
\frac{dR_i}{dt}&= \epsilon (A_i - \gamma R_i+C) \\\nonumber
\frac{dS}{dt}&= \alpha_f + \rho \alpha_0 + \rho D\frac{1}{N}\sum_{i=1}^N \Theta(A_i)-JS,
 \end{align}
 where $i$ is the index of cells changing from 1 to the total number of cells, $N$. The variable $A_i$ and $R_i$ are the internal states of the $i$'th cell and correspond to activator and repressor, respectively. $S$ is the concentration of extracellular cAMP and $I(S)$ is the preprocessing module, $\rho$ is the density of cells, $D$ measures the amount of cAMP released into the environment when a cell spikes, and $J$ is the total degradation rate of  the extracellular cAMP. 
Finally, we have incorporated stochasticity using a Langevin term,  $\eta_i(t)$. In particular, $\eta_i(t)$ is an additive Gaussian white noise term with mean and correlation defined as:
\begin{align}
\left<\eta(t)\right>&=0\\\nonumber
\left<\eta_i(t)\eta_j(t')\right> &=\sigma^2\delta_{ij}\delta(t-t')
\end{align}
The model and corresponding parameters are summarized in  figures \ref{fig:Schematic}A and \ref{fig:Schematic}B. 

Using this model, we can explore a series of questions about how the architecture of the {\it Dictyostelium} signaling circuit within cells affects population-level behaviors. Recent experimental data suggests that the behavior of  {\it Dictyostelium} circuit is well described by the logarithmic preprocessing module and responds to fold changes in extracellular cAMP  \cite{Sgro2014}. This leads to natural questions about the consequences of pre-processing in the {\it Dictyostelium} signaling circuit. In particular, using our model we will examine how {\it Dictyostelium}  exploits the interplay between stochasticity, excitability, and signal processing to control population-level behavior.

\section{Behavior for large degradation rates of extracellular cAMP}
\label{sec:LargeDegradationRate}

\subsection{The quasi-steady-state limit}
In general, the dynamics described by the family of models described by  Eq. \eqref{eqn:Model} are quite complex. For this reason, it is worth considering various limits in which the dynamics simplifies. One such limit that can be realized experimentally is the limit where the extracellular cAMP is degraded quickly compared to the dynamics of the {\it Dictyostelium}  circuit.  This limit can be realized experimentally by changing the flow rate of buffer into the microfluidic device (see Fig. \ref{fig:Schematic}).  In this limit, there exists  a separation of time-scales between external medium and individual oscillators and we can treat the extra-cellular cAMP as a fast variable that is always in a quasi-steady state and set $dS/dt=0$ in  \eqref{eqn:Model}. In this limit, one has
\begin{align}
S\approx{\alpha_f+\rho\alpha_0\over J} + {\rho D\over J}\frac{1}{N}\sum_{i=1}^N \Theta(A_i).
\label{eq:SLargeJ}
\end{align}
For the remainder of this section, we will work within this quasi steady-state approximation. A formal definition of what constitutes large  $J$ will be discussed in section \ref{sec:SmallDegradationRate} where we will give numerical evidence showing that there exist a minimum value $J_m$ above which this approximation is valid.

In this limit, it is helpful to  divide the extracellular cAMP into two terms that reflect the two mechanisms by which extracellular cAMP is produced (see Fig. \ref{fig:Schematic}). First, cells can secrete cAMP at a basal rate $\alpha_0$. We denote the extracellular cAMP produced by this basal leakage,  $S_0$, and in the quasi steady-state approximation this is given by 
\begin{align}
S_0&\equiv{\alpha_f+\rho\alpha_0\over J}.
\label{eqn:Axes1}
\end{align}
where the experimental input flow, $\alpha_f$ is also incorporated into the definition. The second mechanism by which extracellular cAMP is produced is through the release of cAMP into the environment when cells spike. We can parameterize the extracellular cAMP produced by this latter mechanism by 
\begin{eqnarray}
\Delta S  &\equiv& {\rho D\over J },
\label{eqn:Axes2}
\end{eqnarray}
with the total extracellular cAMP produced by spiking given by the expression,
\begin{align}
\Delta S \left<\Theta(A)\right> \equiv  \Delta S\frac{1}{N}\sum_{i=1}^N \Theta(A_i)
\label{squarewave}
\end{align}
To better understand the quantities $S_0$ and $\Delta S$, it is useful to consider an ideal situation where all the cells in a population are perfectly synchronized. In this case,  $\left<\Theta(A)\right>$ will periodically switch between $0$ and $1$. Hence $S$ will behave like a square wave with baseline $S_0$ and amplitude $\Delta S$ (see figure \ref{fig:Schematic}C). Thus, $S_0$ corresponds to the cAMP levels in the troughs and $S_0+\Delta S$ the levels at peaks. These two quantities provide us with a succinct way to represent our model and in the following section and we will use them to produce phase diagrams in the large $J$ regime.  Finally, we note that the square wave form of $S$ is merely a result of our choice of Heaviside function in dynamics of the external medium. Nonetheless, the basic separation of time scales discussed above holds even when the Heaviside function is replaced by a more realistic smooth function. 

\subsection{Phase diagrams for population level oscillations}
\label{sec:PhaseDiagram}

Populations of {\it Dictyostelium} cells can exhibit several qualitatively distinct behaviors depending on the parameters of our model. Cells in a population can oscillate in phase with each other resulting in synchronized, population-level oscillations. We will call this behavior synchronized oscillations (SO). Alternatively, individual cells may oscillate, but the oscillations of the cells are out of phase. In this case, the phase differences between cells prevent the formation of coherent population level oscillations and we call these incoherent oscillations (IO). Finally, even individual cells may not  spike. We will label this behavior No Oscillations (NO). To distinguish between these behaviors,  it is useful to define three order parameters: the coherence, the single-cell firing rate, and population firing rate. Coherence measures how synchronized cells are within a population and is $1$ for a completely synchronized population and $0$ for a fully incoherent one (see Appendix \ref{app:Coherence} for a formal definition). To determine the rate at which a cell $i$ spikes, we count how often the activator variable $A_i$ becomes positive over some averaging time. We then average the firing rate of individual cells over the population. Finally, we normalize the rate so that the single cell firing rate is $1$ for fast oscillations and is $0$ in the absence of spiking (see Appendix \ref{app:FiringRate} for a formal definition). The population firing rate is defined as the firing rate of the average of activator across all cells in the population, $\left< A_i \right>$ and is also normalized to be between 0 and 1. Note that when we calculate the population firing   rate we are measuring whether the average activator over all cells exhibits a spike. If cells are unsynchronized, this average activator $\left< A_i \right>$ will not exhibit any spikes. Thus, population firing rate is a measure of spike counts in a population that fires synchronously.

\begin{figure}[h!]
\centering
\includegraphics[width=\linewidth]{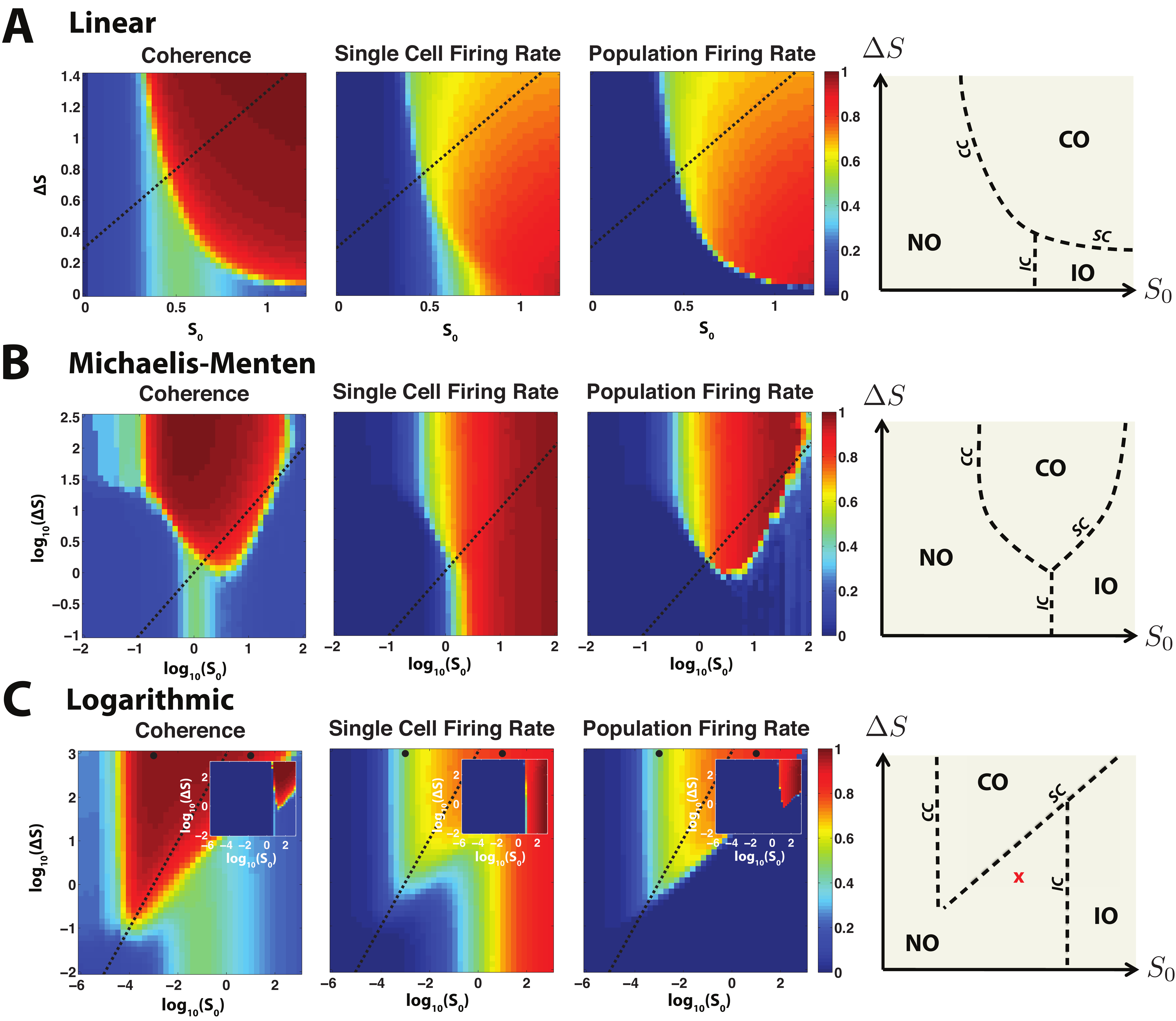}
\caption{\textbf{System Phase Diagram - }{\bf A)} The first three plots from left are phase diagrams of coherence, single cell firing rate and population firing rate as a function of $S_0$ and $\Delta S$, in the large $J$ regime for linear preprocessing. The dashed line corresponds to values of $\Delta S$ and $S_0$ for which $D=2, \alpha_0=1$ and $\alpha_f=0$ with variable $\rho$ and  $J$. Parameters are $J=10, \epsilon = 0.2, \gamma= 0.5, C = 1, \sigma= 0.15, dt= 0.005, t_{wt}=1000, t_{rt}=4000, N= 100$ and $I(S)=S$.
The rightmost plot is a schematic of the phase diagrams marked with different regions. The regions consist of NO: No Oscillation, CO: Coherent Oscillation, IO: Incoherent Oscillation. For easier reference to different transitions the following lines have been introduced: SC: Sensitivity Crossover, IC: Incoherent Crossover, CC: Coherent Crossover 
{\bf B)} Same plots as in (A) with a Michaelis Menten preprocessor. Parameters are same as in (A) with $t_{wt}= 11000$ and $I(S)=\beta S/(S+K_D)$ where $K_D=2.0, \beta=1.5$. The dashed line is plotted for $D=1000, \alpha_0=1000$ and $\alpha_f=0$.
{\bf C)} Same plots as in (A) with logarithmic preprocessor.  The black dots correspond to parameter values chosen in figure \ref{fig:FrequencyIncrease}A. The dashed line is plotted for $D=1000, \alpha_0=1$ and $\alpha_f=0$.
Parameters are same as in (A) with $I(S)=a \ln(S/K+1)$ where $a=0.058, K=10^{-5}$
. Inset is the same plots for a noise level 10 times smaller ($\sigma=0.015$) run for a longer waiting time ($t_{wt}=50000$). 
}
\label{fig:PhaseDiagrams}
\end{figure}

\begin{figure}
\centering
\includegraphics[width=\linewidth]{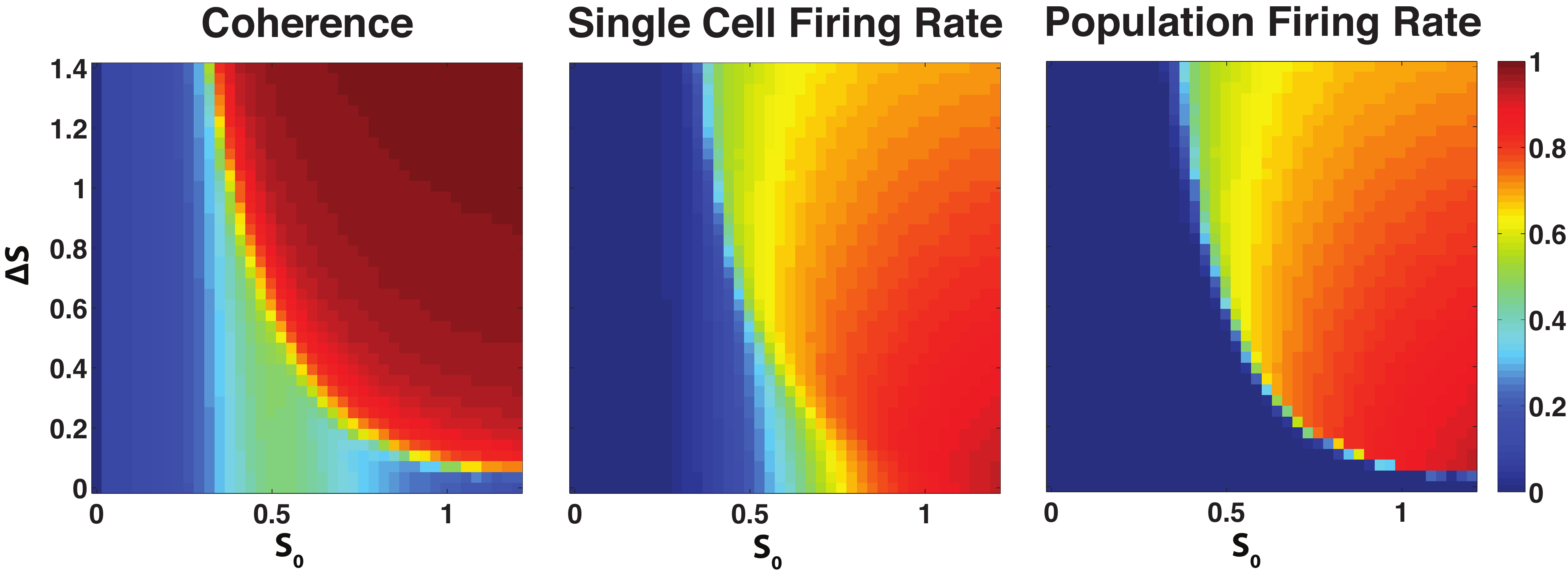}
\caption{ \textbf{SI - Phase Diagram with Large J  -} Same simulation as in figure \ref{fig:PhaseDiagrams}A with larger degradation rate ($J=100$). All the other parameters were kept the same as in figure \ref{fig:PhaseDiagrams}A. 
}
\label{fig:Alpha0DPhaseDiagJ100}
\end{figure}

Using these order parameters, we constructed phase diagrams characterizing the population level behavior for large degradation rates as a function of  $S_0$ and $\Delta S$ (see equations \eqref{eqn:Axes1},\eqref{eqn:Axes2}). We calculated the coherence, single cell firing rate and population firing rate for equation \eqref{eqn:Model} for our three preprocessing modules as a function of $S_0$ and $\Delta S$ (see  Figure \ref{fig:PhaseDiagrams}). Each data point on these phase diagrams corresponds to one simulation of equation \eqref{eqn:Model} for a fixed simulation time (see Appendix \ref{app:ForwardIntegration}) where $J$, $\alpha_0$, and $\rho$ are kept the same for the whole phase diagram and $\alpha_f$ and $D$ are chosen such that the desired $S_0$ and $\Delta S$ are achieved. Finally,  we checked that phase diagram was insensitive to a ten fold increase in the degradation rate $J$ confirming our assumption that the dynamics depend on the parameters only through $\Delta S$ and $S_0$ (see figure \ref{fig:Alpha0DPhaseDiagJ100}). 
                                    
 This phase diagram contains three qualitatively different regions. We have labeled these different regions with NO for No Oscillation, CO for Coherent Oscillation and IO for Incoherent Oscillation. The crossover between these regions, which will be explained below, is shown by dashed lines and is labeled as CC for Coherent Crossover, IC for Incoherent Crossover and SC for Sensitivity Crossover. Note that the boundaries between different regions is approximate and has been achieved simply by a rough thresholding. In reality there is no sharp transition from one `region' to another but instead due to noise this `transition' happens in a continuous way. This is a general feature of noisy bifurcating systems and in practice, depending on the problem under study, a reasonable threshold has to be chosen to separate different qualitative behaviors. As a result the schematic phase diagrams drawn in figure \ref{fig:PhaseDiagrams} are just rough sketches of the qualitative behavior of the system.

\begin{figure}
\centering
\includegraphics[width=\linewidth]{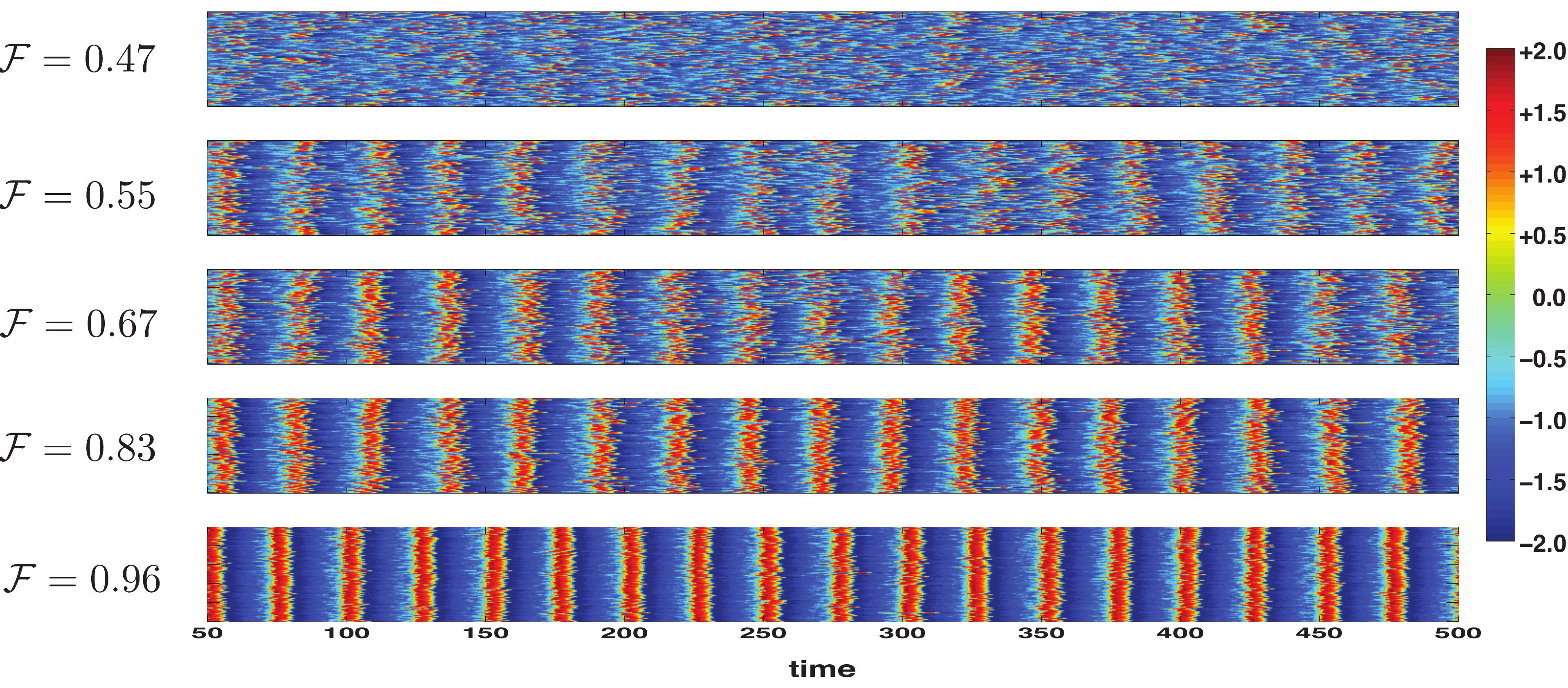}
\caption{ \textbf{SI - Samples for Coherence  - } Raster plots of a population of 100 oscillating cells are shown as a function of time for different levels of coherence $\mathcal{F}$. From this figure it can be seen that $\mathcal{F}\approx0.6$ is a reasonable threshold for separating coherence from incoherence.
}
\label{fig:SamplesForCoherence}
\end{figure}

In the NO region, single cells are silent and oscillations do not occur. As a result, both coherence and single cell firing rate are zero. In this region, the basal level of cAMP, $S_0$, is so small that the cells cannot be excited into oscillation. Note that at all points in the NO region, the parameters are such that individual cells are below the bifurcation to oscillations in the FHN model. However, even below the bifurcation cells occasionally spike due to stochasticity.  In the CO region, cells oscillate coherently. This can be seen by noting that single cell firing rate is nonzero and coherence is close to one. By studying multiple time-courses we found that cell populations with coherence values approximately above 0.6 can be considered coherent (see figure \ref{fig:SamplesForCoherence}). 

In the IO region, single cells oscillate but are unsynchronized. On the phase diagrams these are regions with large values of single cell firing rate (close to $1$) and small values of coherence (approximately less than 0.6). In this region individual cells oscillate and, in each firing, secrete cAMP into the environment. However this change in extracellular cAMP is not enough to excite the cells to synchronize. To understand the reason behind this, we need to look at changes in the input, $I(S)$, that the excitable systems receive. For a population of cells that is oscillating coherently, $S(t)$ can be thought of as a square wave oscillating between $S_0$ and $S_0+\Delta S$ (see Figure \ref{fig:Schematic}C). Then the input of each excitable module within cells can be visualized as a square wave oscillating between $I_0$ and $I_0+\Delta I$ with:
\begin{align}
\label{eqn:DeltaI}
I_0&=I(S_0)\\\nonumber
\Delta I &= I(S_0+\Delta S)-I_0
\end{align}
If changes in $\Delta I$, are smaller than the FHN's sensitivity for signal detection, single cells will instead experience a constant input with no discernable fluctuations and cannot be coherent. For a preprocessor with a monotonically increasing convex functional form, $I(S)$ (with $I(0)=0$) such loss of coherence may happen due to very small $\Delta S$ or very large $S_0$. 

Our phase diagrams exhibit a number of crossovers between the three qualitative behaviors outlined above. The Incoherent Crossover (IC) separates regions with  no oscillations from those where cells oscillate incoherently. This transition occurs when $\Delta S$ is not large enough to produce any discernible changes in the external medium. As a result each individual cell goes through this crossover as if it was alone and not communicating with other cells. For these uncoupled cells, as $S_0$ is increased the system gets closer to bifurcation and fires more often. Figure \ref{fig:SingleCellFreq} shows this increase in firing rate for a single cell corresponding to $\Delta S=0$.

There is also a crossover from the no oscillation region to coherent population level oscillations. We have labeled this the Coherent Crossover (CC). Here, as $S_0$ is increased, individual cells become more likely to spontaneously spike. These spontaneous spikes happen because, given a monotonically increasing function $I(S)$, for larger $S_0$ the excitable system's input will be closer to bifurcation point, causing the system to become more excitable. As a result, noise can more often kick the system out of its stable fixed point, leading to a spike commonly referred to as an accommodation spike \cite{Izhikevich2007}. If $\Delta S$ is large enough, a single (or few) cell's spike will be enough to cause a change in the external medium that can be sensed by other cells. The other cells will then get excited with a small delay, creating the effect of a synchronized spike. Because FHN has a refractory period, for sometime no cell will spike, until the effect is repeated again. The overall behavior in this way seems similar to coherent oscillations, but is in reality periodic synchronized noise-driven spikes that are happening way below the system's bifurcation point. To show that this effect is noise-dependent we decreased noise by an order of magnitude for the system with logarithmic preprocessor and plotted the results (inset of \ref{fig:PhaseDiagrams}C). We observed that CC shifted to the same value of $S_0$ as IC, indicating that the `knee' shaped region (intersection of CC and SC) emerges due to noise.

Finally,  Sensitivity Crossover (SC) separates regions with coherent oscillation from those with incoherent or no oscillations.  As one crosses the SC line, cells lose their ability to detect the changes in the external medium. Each excitable system has a response threshold and cannot respond to abrupt changes in its input if they are below this threshold. In our model this can occur either because $\Delta S$ is very small or due to the nonlinear form of preprocessor. The former case is a manifestation of the fact that for very small changes in the external medium, cells do not have any means of communication. However the latter case requires some further explanation. For two of the preprocessors used in our simulations (i.e. Michaelis-Menten and logarithmic) the function $I(S)$ was chosen to be concave and monotonically increasing. This means that, for a fixed $\Delta S$, as $S_0$ is increased $\Delta I$ in equation \eqref{eqn:DeltaI} decreases. Once $\Delta I$ goes below the detection sensitivity of excitable modules, coherence will be lost. Note that since increasing $S_0$ and/or decreasing $\Delta S$ lead to decrease of $\Delta I$, for larger values of $\Delta S$ a larger value of $S_0$ is required to take the system from coherence to incoherence (assuming that sensitivity of excitable system is roughly independent of baseline $I_0$). This is why in figure \ref{fig:Schematic}B,C the slope of SC is positive.

An interesting observation is that the preprocessing module into the excitable system can dramatically change the phase diagram of the cellular population. This suggests that it is possible to have different population level behaviors from the same underlying excitable system by changing the upstream components of a network. We now briefly discuss the differences in phase diagrams arising from the choice of pre-processing modules. The first row in figure \ref{fig:PhaseDiagrams}A shows the phase diagrams for a linear preprocessor. As can be seen in the schematic (last column), the curve for SC is almost flat (with a slight downward slope), a signature that single cell's sensitivity to changes in the external medium is almost independent of the baseline $S_0$. However inclusion of a preprocessing module completely changes this effect. Figure \ref{fig:PhaseDiagrams}B and figure \ref{fig:PhaseDiagrams}C show the results for a Michaelis-Menten and logarithmic preprocessor respectively. Note that in both cases SC has a dramatic positive slope. This is due to the concave monotonically increasing nature of the preprocessors chosen. 

It is interesting to note that for the logarithmic preprocessor there is an extra `knee' (where CC and SC intersect) that does not exist when Michaelis-Menten is used. A behavior reminiscent of this subregion has been observed experimentally \cite{Sgro2014} where increasing $S_0$ (by changing input flow) for a synchronously oscillating population destroys the oscillations, whereas further increase leads to incoherent oscillations. This suggests that the system is tuned to this corner. Interestingly, in this region of phase diagram $S_0$ and $\Delta S$ take the smallest possible values that can lead to coherent oscillations. Since $S_0$ and $\Delta S$ both correspond to production of cAMP by the cell, it seems reasonable from an evolutionary point of view if the system is fine-tuned to minimize its cAMP production while achieving the coherent oscillations. Experimentally, it is possible to move horizontally along this phase diagram by changing $\alpha_f$ and move laterally by changing $\rho$ and $J$. One interesting prediction of this phase diagrammatic approach is that for a coherently oscillating population of cells with a certain cell density increasing degradation rate, $J$ should decrease the minimum cAMP flow ($\alpha_f$) required to destroy the oscillations, an observation that has been confirmed experimentally \cite{Sgro2014}.

\begin{figure}
\centering
\includegraphics[width=\linewidth]{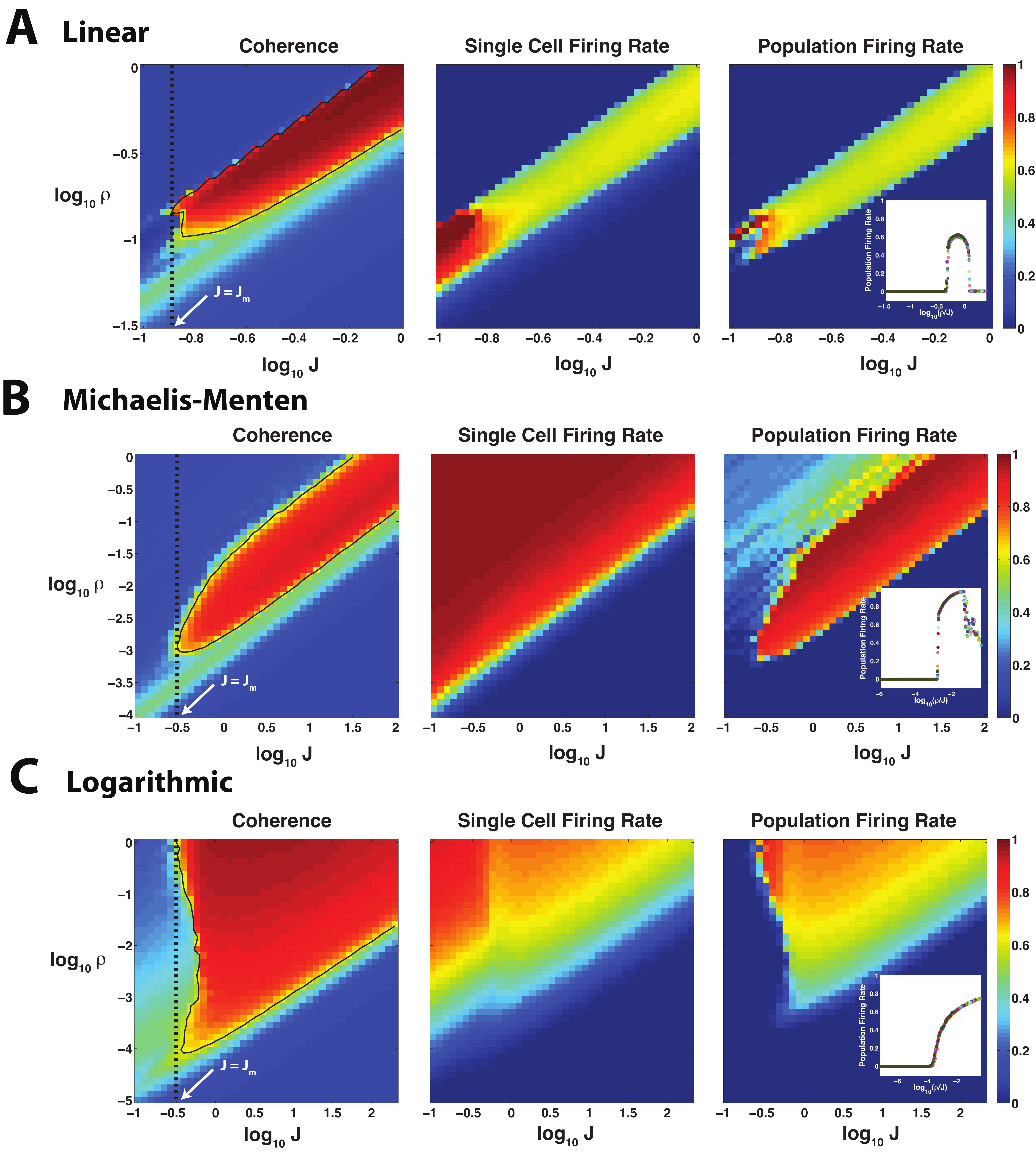}
\caption{\textbf{Effect of $\rho$ and $J$ - } {\bf A)} Plot of coherence, single cell firing rate and population firing rate for different values of $\rho$ and $J$ with linear preprocessor. Parameters same as in figure \ref{fig:PhaseDiagrams}A with $\alpha_0=1, D=2, \alpha_f=0$ corresponding to dashed line in figure \ref{fig:PhaseDiagrams}A. The black curve in the coherence graph is where coherence is equal to $0.6$, marking an approximate boundary for crossover between coherence and incoherence. The dashed line is the leftmost line with constant $J$ that intersects with the balck curve. We have called the value of $J$ on this line $J_m$. The inset is population firing rate as a function of $\rho/J$, showing a data collapse for which data points are taken from the population firing rate heat map. To avoid effects of small degradation rate only values with $J>3J_m$ are plotted in the inset. {\bf B)} Same plot as in (A) with Michaelis-Menten preprocessing. Parameters same as in \ref{fig:PhaseDiagrams}B with  $\alpha_0=1000, D=1000$ corresponding to dashed line in figure \ref{fig:PhaseDiagrams}B. The inset is plotted for $J>10J_m$. {\bf C)}  Same plot as in (A) with logarithmic preprocessing. Parameters same as in \ref{fig:PhaseDiagrams}C with  $\alpha_0=1, D=1000$ corresponding to dashed line in figure \ref{fig:PhaseDiagrams}C. The inset is plotted for $J>10J_m$. 
}
\label{fig:rhoJHeatMap}
\end{figure}

Experimentally it is possible to change both $\rho$ and $J$. Gregor {\it et al} \cite{Gregor2010} measured population firing rate for different values of $\rho$ and $J$. They showed that there is a data collapse of the population firing rate  as a function of $\rho/J$. To understand this behavior, we made phase diagrams as a function of $\rho$ and $J$ (Figure \ref{fig:rhoJHeatMap}). The insets show that for large degradation rates, this data collapse occurs for all choices of pre-processing modules. The underlying reason for this is that, as discussed above, in this limit the population dynamics depends on the external medium only through $S_0$ and $\Delta S$. Both these quantities depend on $\rho$ and $J$ through the combination ${\rho \over J}$ (see Eq \ref{eqn:Axes1} and Eq. \ref{eqn:Axes2}).

\subsection{Frequency increase as a function of density}
\label{sec:FrequencyIncrease}
Our model also suggests a mechanism for cell populations to tune their frequency in response to steps of cAMP. An example of a time-course simulation of this behavior is shown in figure \ref{fig:FrequencyIncrease}A. In this figure a step of external cAMP is flowed into a population of coherently oscillating cells, leading to an increase in the frequency of oscillations. This frequency increase suggests that populations can tune their frequency by modulating the cAMP secretion and excretion rates. 

To explain the underlying reason for the frequency increase, it is useful to consider  the extreme case of a perfectly synchronized oscillating population.  For this case the extracellular cAMP concentration, $S(t)$, will be a square wave that oscillates between $S_0$ and $S_0+\Delta S$ (see  figure \ref{fig:Schematic}C). As a result, the input to the FHN module will be a square wave oscillating between $I_0$ and $I_0+\Delta I$ (see equation \eqref{eqn:DeltaI}). Thus, the dynamics of individual cells can be thought of as an FHN that periodically switches between two cubic nullclines, corresponding to the inputs $I_0$ and $I_0+ \Delta I$. A schematic of the phase portrait of this system is shown in figure \ref{fig:FrequencyIncrease}B for two different values of $\Delta I$. As can be seen from this figure, a decrease in $\Delta I$ decreases the distance between the two cubic nullclines and leads to a shorter trajectory, hence larger frequency. So any mechanism that can decrease $\Delta I$ can lead to an increase in frequency. One such mechanism is by exploiting the nonlinearity of the preprocessing module. Note that in our example $\alpha_f$ is being increased while other parameters of the system are kept constant. This is equivalent to increasing $S_0$ while keeping $\Delta S$ constant. Since $I(S)$ is a monotonically increasing concave function, given a constant value of $\Delta S$ an increase in $S_0$ will lead to a decrease in $\Delta I$ (see figure \ref{fig:FrequencyIncrease}). And this, in turn, leads to an increase in frequency. In practice, there are two other mechanisms that also contribute to frequency increase. However, they are not multicellular effects and happen independent of preprocessing module. The interested reader can refer to Appendix \ref{app:FrequencyIncrease} for a detailed description.

Note that to observe this behavior we have tuned the parameters of the system to a different point (black dots in figure \ref{fig:PhaseDiagrams}C) than what has been used in the rest of the paper. This change of parameters was necessary to ensure that initially the cells were not oscillating at maximum frequency and yet would stay synchronized as $\alpha_f$ was increased. As a result it may not be possible to observe this behavior in wildtype {\it Dictyostelium} cells. However, it is of interest as a theoretical mechanism for frequency tuning and has the potential to be observed in {\it Dictyostelium} mutants or be implemented in synthetic biological networks of other organisms.

\section{Small Degradation Rate and Bistability}
\label{sec:SmallDegradationRate}

Thus far, we have studied the behavior of our model in the large $J$ regime. In this section, we will instead focus on the regime where this parameter is small. For small values of $J$, the dynamics of the external medium becomes too slow to follow the oscillations of single cells. As a result, cells become unsynchronized. A behavior somewhat similar has been termed `dynamic death' by Schwab {\it et al.} \cite{Schwab2012}. These authors studied a population of oscillators coupled to an external medium and observed incoherence due to inability of external signal to follow individual oscillators. In their system, the cause of incoherence was slow response of the external medium to inputs rather than slow degradation. However, in both cases,  the underlying reason for the loss of coherence is the slow dynamics of external signal.

\begin{figure}
\centering
\includegraphics[width=\linewidth]{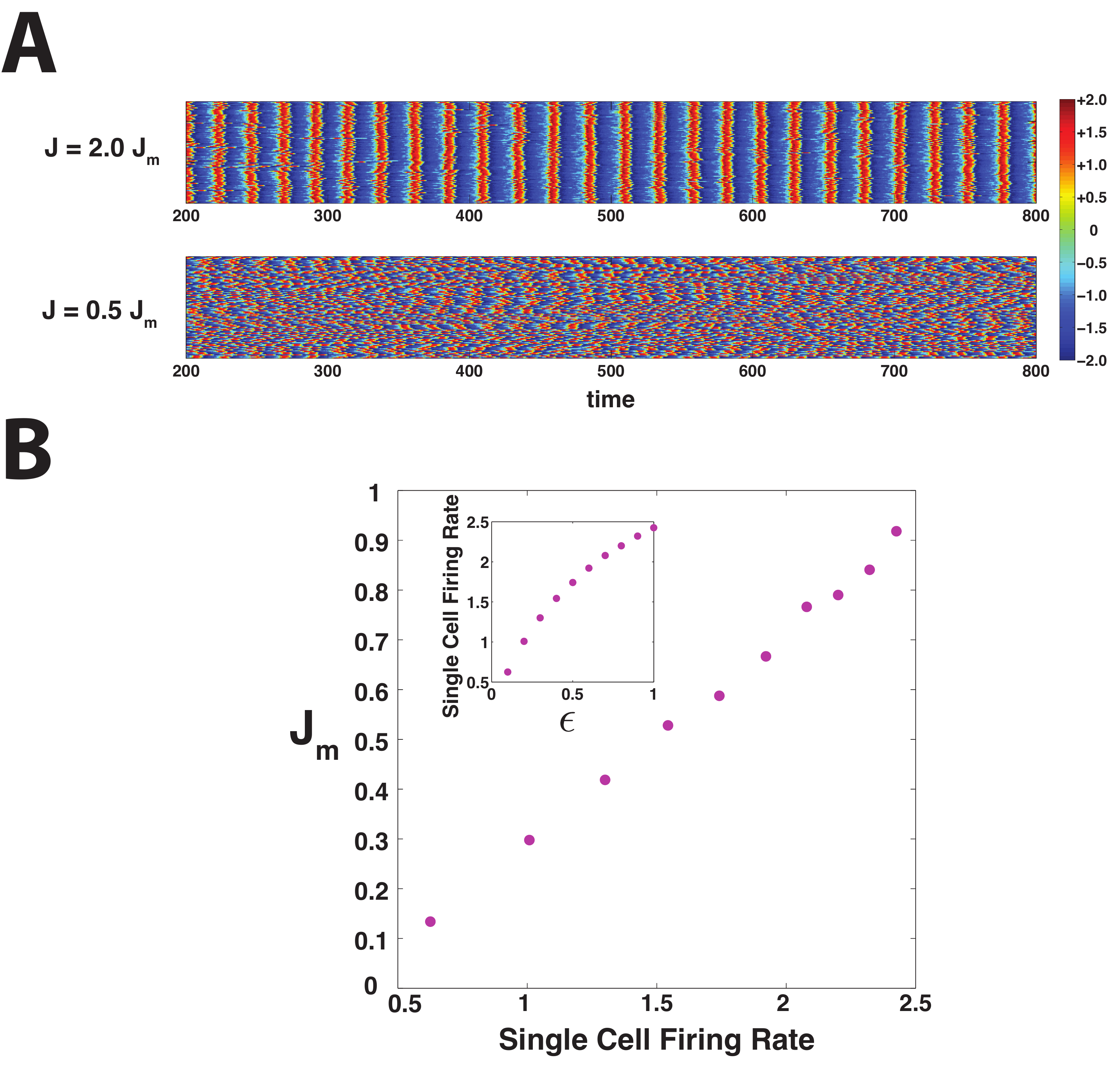}
\caption{ \textbf{Small $J$ Regime  - }  {\bf A)} A raster plot of $A$ as a function of time for degradation rate ($J$) greater and smaller than $J_m$ showing a crossover to incoherence as $J$ is decreased. Each row in the raster plot corresponds to time-course of activator of one cell within the population. Parameters same as in figure \ref{fig:PhaseDiagrams}C with $\rho=1, \alpha_f=0$ {\bf B)} Plot of $J_m$ as  a function of single cell firing rate. Firing rate is chnaged by changing $\epsilon$ while keeping all other parameters same as in part (A). The inset shows how the single cell firing rate changes as a function of $\epsilon$. Parameters same as in part A.  
}
\label{fig:SmallJ}
\end{figure}

We can numerically define a minimum degradation rate, $J_m$, below which the dynamics of the external medium are too slow to sustain population level oscillations. To do so, we identified the boundary separating the region of coherence from incoherence by thresholding the coherence at approximately $0.6$. This boundary is indicated by the  black curves in the coherence plots in the  first column of Fig. \ref{fig:rhoJHeatMap}. We call the smallest value of $J$ on this curve the minimum degradation rate, $J_m$. Figure \ref{fig:SmallJ}A shows a raster plot of the oscillations for $J=2 J_m$ and $J=0.5 J_m$, with all other parameters fixed. Notice that decreasing $J$ below $J_m$ completely destroys the ability to have synchronized population-level oscillations. Finally, it is worth  emphasizing that due to the stochastic nature of our system, there is no sharp transition from coherence to incoherence at $J_m$. Rather,  $J_m$ serves as a crude, but effective scale, for understanding external medium dynamics. 

\begin{figure}
\centering
\includegraphics[width=0.6\linewidth]{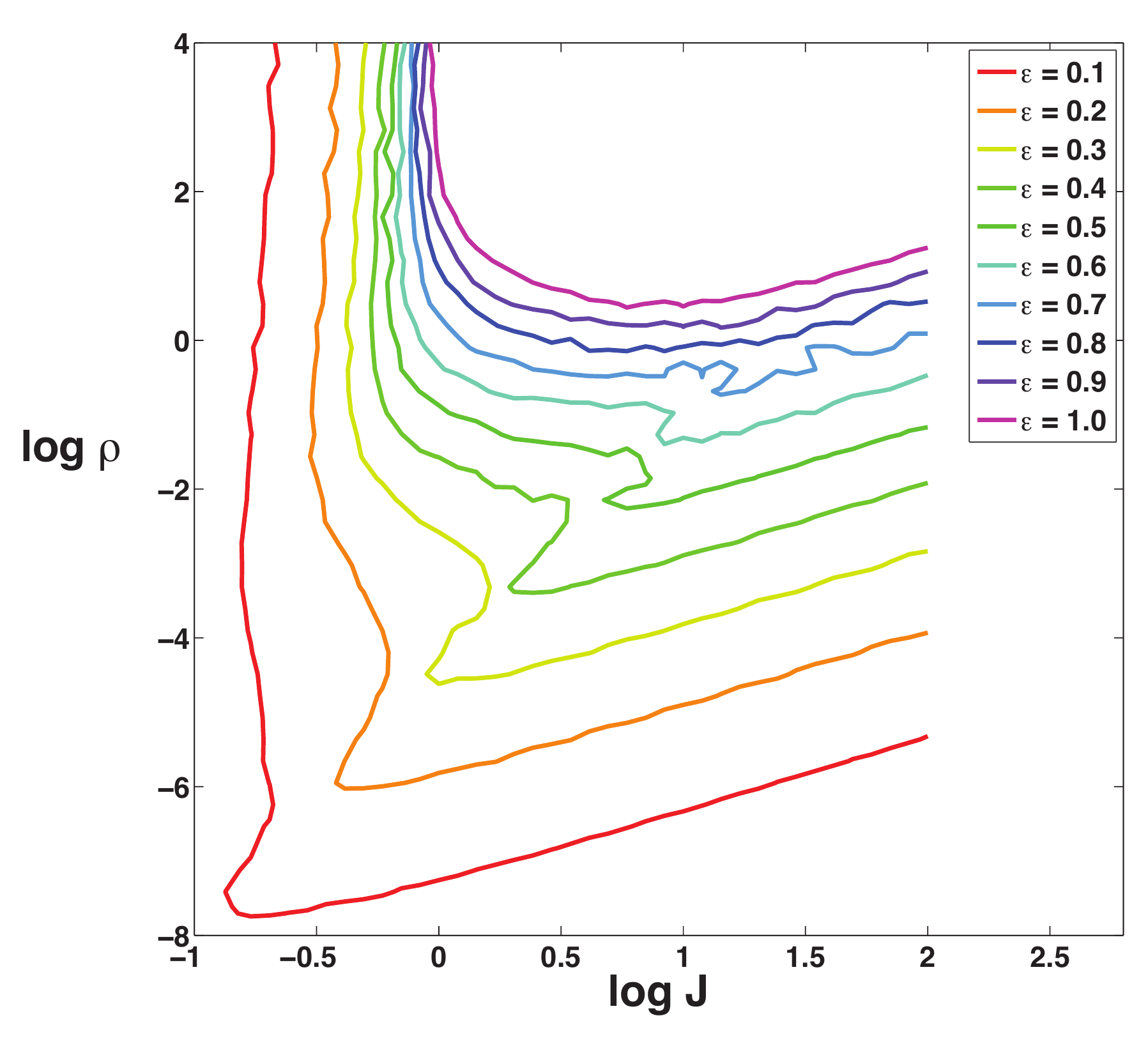}
\caption{ \textbf{SI - Boundaries  - } Boundary of coherence heat map is plotted for different values of $\epsilon$. Each curve is plotted similar to the black curve in figure \ref{fig:rhoJHeatMap}C. Parameters same as in figure \ref{fig:rhoJHeatMap}C. 
}
\label{fig:Boundaries}
\end{figure}

To better understand $J_m$,  we asked how it scaled with the  cell period in the underlying FHN model. Since $J_m$ is a measure of when the external signaling dynamics are slow compared to the signaling dynamics of individual cells, we hypothesized that $J_m$ would scale with the frequency of oscillations in the underlying FHN model. To test this hypothesis, we changed single cell frequencies by changing $\epsilon$ in equation \ref{eqn:Model}. We then determined $J_m$ in each case by finding the boundary between coherence and incoherence (see figure \ref{fig:Boundaries}). The results are shown in figure \ref{fig:SmallJ}B. As postulated, we find that increasing single cell firing rate leads to a higher value of $J_m$.  These results numerically confirm our basic intuition that $J_m$ is a measure of when the  external signal response is much slower than the time scale on which individual cells respond to stimuli.

\begin{figure}
\centering
\includegraphics[width=\linewidth]{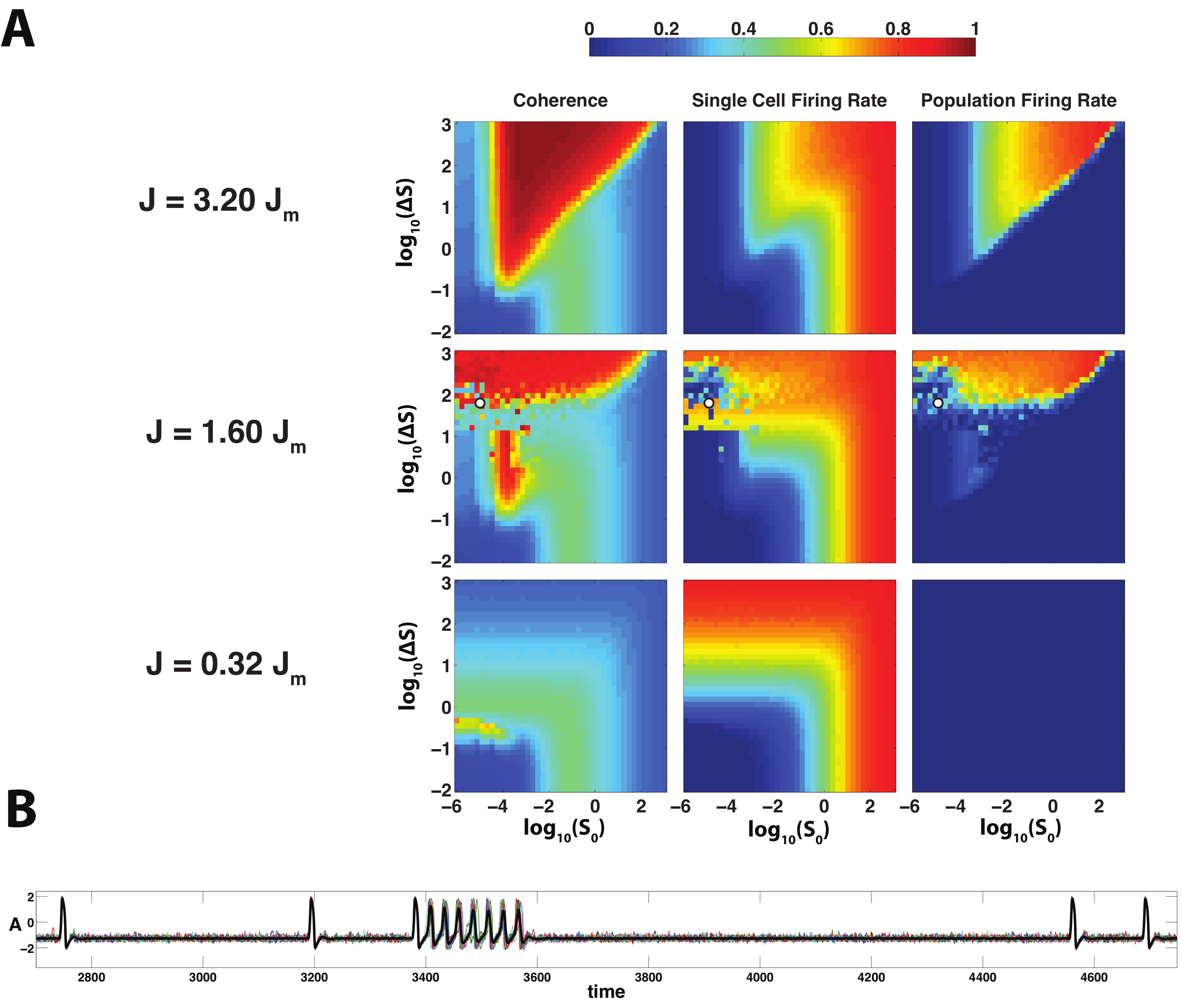}
\caption{ \textbf{Bistability at Crossover to Small $J$ - } {\bf A)} Plot of coherence, single cell firing rate and population firing rate as a function of of $\log_{10}(S_0)$ and $\log_{10}(\Delta S)$ for three different values of degradation rate $J$. The white circle corresponds to one point on the phase diagram with $J=1.6J_m$ for which a time-course is shown in figure B. Parameters same as in figure \ref{fig:PhaseDiagrams}C {\bf B)} A section of the time course of the system is shown with parameters chosen corresponding to the white circle in figure A (middle row). Each thin curve with a different color corresponds to time course of the activator of one cell. For presentation purposes nnly 10 cells are shown (picked at random). The black curve is the time-course of the average of activators of all cells.  Parameters same as in part C with $J=0.5, \rho=1, D= 10^{1.8}J, \alpha_0=10^{-5.1}J$  
}
\label{fig:BistabilityAll}
\end{figure}

In section \ref{sec:PhaseDiagram} we studied the system in the $J\gg J_m$ regime. Here, we re-examine the phase diagrams changes in the opposite limit when  $J$ is decreased below $J_m$. To this end, we produced a set of phase diagrams with different values of $J$. Figure \ref{fig:BistabilityAll} shows three representative phase diagrams showing this crossover. Notice that the phase diagram above $J_m$ at $J=3.2J_m$ is very similar to \ref{fig:PhaseDiagrams}C; however, decreasing $J$ below $J_m$ to $J_m=0.32J_m$ creates a completely incoherent population in which single cells can oscillate in regions that previously contained no oscillations (NO). This is likely due to the fact that once a cell fires, the secreted cAMP takes a very long time to be degraded. During this time, other cells can spontaneously fire. These spiking events are incoherent but still give rise to elevated levels of external cAMP.

More interestingly the transition from the behavior at large degradation rates ($J > J_m$) to small degradation rates ($J<J_m$) happens through an intermediate state with many peculiar data points (the middle row in figure \ref{fig:BistabilityAll}A). To ensure that these peculiarities are not simulation artifacts we looked at some of them in more detail. Figure \ref{fig:BistabilityAll}B is a time-course of the whole population for the point corresponding to the white circle on \ref{fig:BistabilityAll}A. Note that the time-course is exhibiting a burst-like behavior. The system is in a low frequency state for some period of time, then stochastically switches to a  high frequency state and after a few cycles switches back to original state. Interestingly, it remains coherent during the whole process. At this point we do not have a conclusive theory as to why this bistability happens. However, a similar behavior had been reported by Schwab {\it et al.} \cite{Schwab2012a} where a population of phase oscillators coupled to an external medium exhibited bistability as mean oscillator frequency was increased. We suspect that a similar mechanism is also in effect here.

\section{Spatial Extension of model produces spiral waves}
\label{sec:SpatialModel}
As a final step in our modeling approach, we extended equation \ref{eqn:Model} to model dense population of {\it Dictyostelium} cells. Here, we restrict ourselves to discussing the biologically realistic case of a logarithmic preprocessing module  $I(S)={a \ln\left( 1+{S\over K}\right)}$, though similar results were obtained for other pre-processing modules. To model dense populations, we treat the activator, $A(x,y)$,  repressor, $R(x,y)$, and extracellular cAMP, $S(x,y)$ as a function of the spatial coordinates $x,y$. Furthermore, we explicitly model the diffusion of the extracellular cAMP.  This gives rise to a set of  reaction-diffusion equations of the form:

\begin{align}
\label{eqn:SpatialModel}
\frac {dA}{ dt} &= A-\frac{1}{ 3}{A}^3-R+I(S) ,\\\nonumber  
\frac{dR}{dt}&= \epsilon \left(A - \gamma R+C\right) \\\nonumber
\frac{dS}{dt}&=\rho \alpha_0 + \rho D\Theta(A)-JS+\nabla^2S
\end{align}

\begin{figure}
\centering
\includegraphics[width=\linewidth]{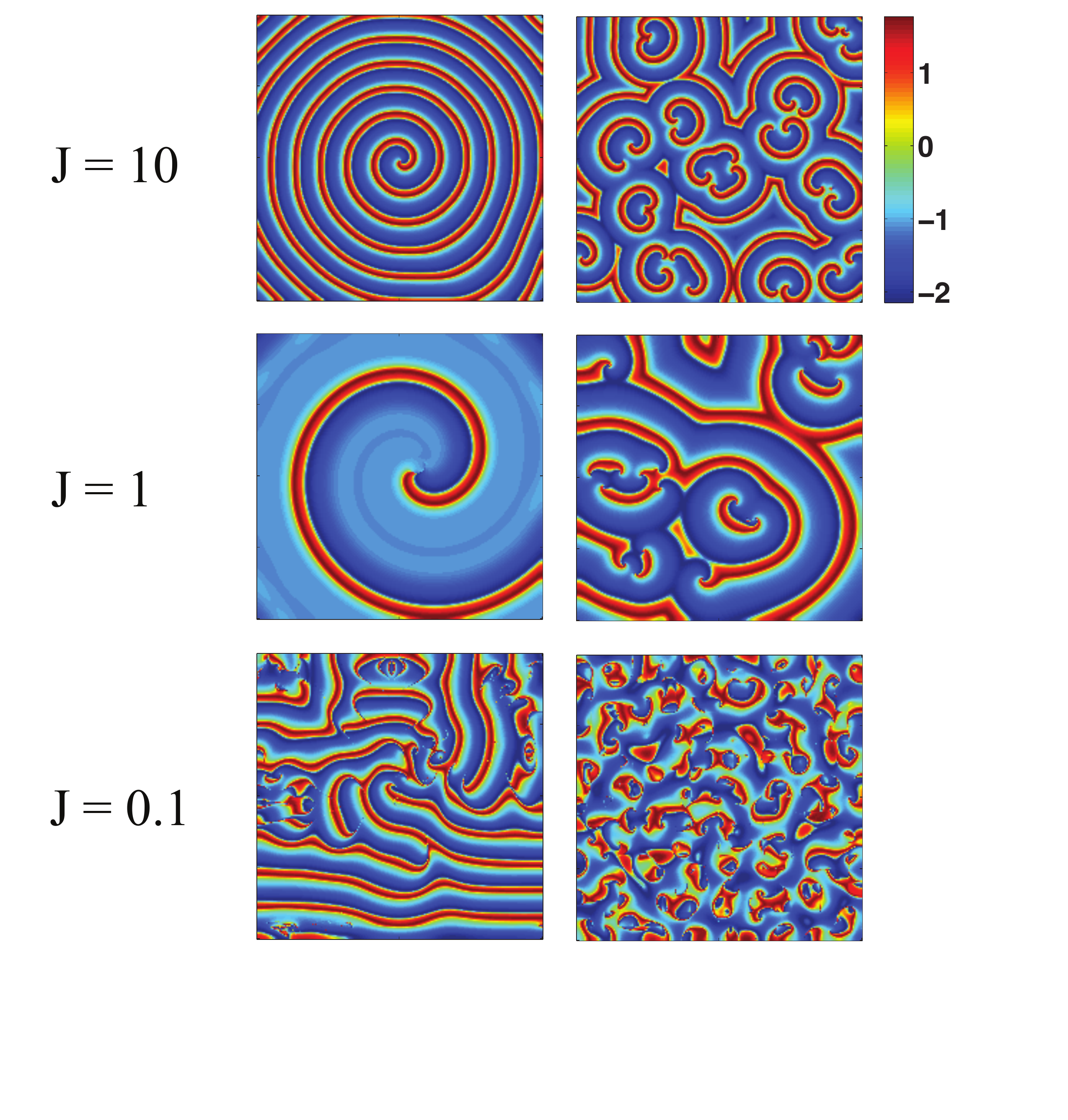}
\caption{ \textbf{ Spatial Simulations -} Simulation results of spatially extended model at different values of $J$. The colors shown represent different levels of $A$. Each row corresponds to a different value of $J$ and $\rho$ such that $\rho/J$ remains the same. The left column corresponds to initial conditions chosen from a $2 \times2$ coarse grid of random values that is overlayed on the simulation box (see Appendix \ref{app:ReactionDiffusion}). And the right column shows the same simulation with initial conditions set on a $20 \times20$ coarse grid. Parameters were kept the same as in \ref{fig:rhoJHeatMap}C with $\rho=0.1J$. Simulations were done on a $100\times100$ box with grid spacing $\Delta x=0.5$ and time steps according to $dt={\Delta x^2\over 8}$. 
}
\label{fig:Spatial}
\end{figure}

For simplicity we have not included the noise term, $\eta$, and the input cAMP flow, $\alpha_f$. Furthermore, diffusion coefficient has been set to 1 by absorbing it into the spatial coordinate. We simulated these equations using no-flow boundary conditions and random initial conditions. Figure \ref{fig:Spatial} shows a snapshot of activator, $A$, over the whole space. The left column shows the results with initial conditions chosen such that at most one spiral forms (see Appendix \ref{app:ReactionDiffusion}). Note that a spiral wave is clearly formed at large values of degradation rate ($J=10$). However, decreasing this parameter while keeping $\rho/J$ constant leads to complete disappearance of the spiral pattern. The right column in figure \ref{fig:Spatial} shows the same results with initial conditions that lead to multiple spirals (see Appendix \ref{app:ReactionDiffusion}). In this case, a similar disappearance of spiral waves is observed as degradation rate of cAMP is decreased. Disappearance of spiral waves has been observed in {\it RegA} mutants\cite{Sawai2005}. Since {\it RegA} intracellularly degrades cAMP, it can be thought of as one contributing factor to degradation rate $J$ in our  simplified model. As a result, knocking out this gene could decrease $J$ and have an adverse effect on spiral formation. In this regard, this simple extension of our model is compatible with experiments.

Besides models of {\it Dictyostelium discoideum}, spiral patterns in excitable media have been observed in many other contexts such as cardiac arrhythmia, neural networks and BZ reactions. In this regard, emergence of spiral patterns in a diffusive excitable medium is not new. However, in the context of {\it Dictyostelium discoideum}, a key difference between our model and previous models such as the one proposed by Aranson {\it et al} \cite{Aranson1996} is that in our model only the external medium $S$ can diffuse. Previous models made the biologically unrealistic assumption that the intracellular variables could diffuse and the external medium did not.

\section{Discussion}

During starvation, {\it Dictyostelium discoideum} cells secrete periodic spikes of cAMP in response to extracellular cAMP levels and communicate by propagating waves of cAMP across space. We modeled this behavior using a multi-scale modeling approach. We constructed a family of dynamical models  that increased in complexity. We started by modeling isolated cells. We then extended to the model to understand  spatially-homogenous multicellular populations. Finally, we included the effects of space and diffusion. In our approach, we treated individual cells as noisy excitable systems that receive their input from a preprocessing module which responds to external cAMP concentrations. We coupled these cells through an external medium and studied their oscillations and coherence through phase diagrams. These phase diagrams provided us with a succinct, interpretable representation of our model. Using these diagrams, we found that the complex interplay of multicellularity, stochasticity and signal processing gives rise to a diverse range of phenomena that have been observed experimentally. By including space into this model we were able to produce spiral patterns and study them in different regimes. 


Using phase diagrams, we showed that the crossover from silence to coherent oscillations is noise-driven. In this process, some cells randomly   fire, leading to the sudden secretion of cAMP and an increase in the external cAMP levels. This change in extracellular cAMP levels induces other cells in the population to spike, resulting in synchronized oscillations across the population. This behavior emerges from the complex interplay of cellular communication and stochasticity. In this process, each population-level spike consists of early spikers and late spikers, where the former drives the latter. This behavior is reminiscent of 'pacemaker' cells which are hypothesized as driving forces for synchronization and pattern formation. But unlike traditional models, in our model no cell is intrinsically a pacemaker. Instead,  early spikers  are picked at random. Thus, noise is crucial to the observed dynamical behavior of cellular populations.

To explore the effect of preprocessor we studied a family of models with different preprocessing modules. We found that the choice of a nonlinear function as the preprocessor leads to a new crossover from coherent oscillations to incoherent oscillations that is non-existent if a linear preprocessor was used. Furthermore, we find that the choice of preprocessors can lead to different responses to noise, with distinct signatures that can be inferred from experimental multicellular data. This allows us to confirm that {\it Dictyostelium} cells use a logarithmic preprocessor, a claim that has been suggested based on independent single cell experiments \cite{Sgro2014}. 

We encountered several interesting behaviors in our model that have implications for other coupled oscillator systems.  For example, we found that the  nonlinearities in the preprocessor can lead to a mechanism for populations of oscillators to change their frequency. Furthermore we found that  slowing the dynamics of the external medium leads to incoherent oscillations. This behavior has been termed `dynamic death' for coupled oscillators \cite{Schwab2012,Schwab2012a}, and we find that it occurs through a bistable state. Furthermore, in the spatial extension of our model, we observe a similar loss of spiral patterns due to slow dynamics of the medium. This suggests that the concept of dynamic death can be extended to spatially heterogeneous populations.

Synchronization and formation of spiral waves provides a spatial cue for {\it Dictyostelium} cells, which guides them toward a common aggregation center. As a result, dynamic death can be undesirable for a population's survival. It is well known that in wildtype cells phosphodiesterases (PDE) are constantly secreted intra- and extracellularly to degrade cAMP. We suspect that this mechanism may have evolved to avoid incoherence due to dynamic death. 


Despite the descriptive and predictive success of our simple model \cite{Sgro2014} it misses several points that could be the subject of future works. For example, we have treated the preprocessor as a static module. However, a more complete model that describes adaptation needs to include the dynamics of this module. Models that contain a feedforward network \cite{Takeda2012, Wang2012, Xiong2010} seem to be good candidates for this purpose. Furthermore, we have ignored the effect of noise in our spatially extended model. It would be interesting to find how noise can affect the random formation of spiral patterns and their stability and explore to what extent a spatially extended model is amenable to the phase-diagrammatic approach proposed here. Finally, it would be interesting to study our model through analytical approaches such as Fokker-Planck equations \cite{Acebron2004,Lindner2004} and explore the possibility of new phases of behavior that have been neglected in our study.

\section{Acknowledgement} We would like to thank Charles Fisher and Alex Lang for useful comments. PM and JN were supported by NIH Grants K25GM086909 (to P.M.). DJS were supported by NIH K25 GM098875-02 and NSF PHY-0957573. The authors were partially supported by NIH Grant K25 GM098875. This work was supported by NIH Grants P50 GM071508 and R01 GM098407, by NSF-DMR 0819860, by an NIH NRSA fellowship (AES), and by Searle Scholar Award 10-SSP-274 (TG).

\appendix
\section{Forward Integration}
\label{app:ForwardIntegration}
In all simulations stochastic differential equations (equation \eqref{eqn:Model}) have been solved using Euler-Maruyama method. The time-step throughout the paper has been $dt=0.005$ unless explicitly stated otherwise. Through trial and error we found that larger time-steps lead to unstable solutions in large parameter regimes and smaller time-steps did not lead to different results. As a result we believe that our choice of time-step produces reliable results. The simulations were started from random initial conditions with $A_i(t=0)$ and $R_i(t=0)$ independently chosen from a Gaussian distribution with mean $0$ and standard deviation $2$ and $S(t=0)$ was set to zero. Although these initial conditions are random and independent for different cells, there still may be correlations between them, meaning that the cells may be partially in phase. To avoid such correlations affecting coherence among cells, we ran each simulation for some waiting time $t_{wt}$ and then continued the simulation for an extra run time $t_{rt}$. The results during the run time are what is shown throughout the paper, while the results during the waiting time were discarded. We found that each simulation required a different amount of waiting time. This was especially dramatic for the case with a very small noise (the inset in figure \ref{fig:PhaseDiagrams}C) where an extremely long waiting time was required. To determine the proper waiting time, we ran each simulation for multiple waiting times and compared the results. Usually when waiting time was too short we could see patterns of `bleeding' in the phase diagram that could be avoided at longer waiting times. By comparing these results in each figure we established a waiting time $t_{wt}$ during which the system could 'forget' its initial conditions.

\section{Firing Rate}
\label{app:FiringRate}
To find the firing rate, $\mathcal{R}(A)$, of a signal $A(t)$ we thresholded the signal compared to zero and counted the number of positive `islands'. By positive 'islands' we mean continuous positive intervals (i.e. $A(t)>0$) that are flanked by negative intervals (i.e. $A(t)\le0$). Such a definition would produce a correct measure of firing rate, if the signal was smooth. However, due to the noisy nature of the simulations, spurious small islands may form, which could be mistakenly counted as extra peaks. To avoid these undesirable counts we will filter out any small `islands'. The procedure is as follows:

We first threshold the signal by defining $B(t)$:
\begin{align}
B(t)=
\left\{
\begin{array}{l l}
    1 & \quad A(t)>0\\
    0 & \quad \text{otherwise}
  \end{array}
 \right.
\end{align}
To get rid of any noise-induced small 'islands' in $B(t)$ we pass it through a low-pass filter. This is done by convolving the signal with:
\begin{align}
H(t)=
\left\{
\begin{array}{l l}
    1 & \quad 0\le t \le \tau\\
    0 & \quad \text{otherwise}
  \end{array}
 \right.
\end{align}
where in all simulations $\tau=1$. To ensure that real peaks are not filtered out, this time-scale is chosen much larger than a typical spurious island width, but much smaller than any real peak width that we ever observed in our simulations. The result of the convolution is then thresholded again to give 
\begin{align}
L(t)=
\left\{
\begin{array}{l l}
    1 & \quad B(t)*H(t)>0\\
    0 & \quad \text{otherwise}
  \end{array}
 \right.
\end{align}

where star stands for convolution. We then count the number of positive `islands' in $L(t)$ which correspond to the number of peaks in $A(t)$. The result is then divided by total time to give the value for firing rate, $\mathcal{R}(A)$. We tested this method on several signals and it was in perfect agreement with counts done manually.

%
\section{Coherence}
\label{app:Coherence}
We defined a measure of coherence among a population of oscillators, $\mathcal{F}$, by treating each cell as a phase oscillator. This was done by treating variables $(A,R)$ of each oscillator as Cartesian coordinates and transforming them into polar coordinates. We then adopt the same definition for coherence used by Kuramoto \cite{Acebron2005}. The definition is such that for a perfectly incoherent system $\mathcal{F}=0$ and for a perfectly coherent system $\mathcal{F}=1$. The mathematical definition of this quantity is as follows:
\begin{align}
A_0&=\int_{t_{wt}}^{t_{wt}+t_{rt} }dt{1\over N}\sum_{k=1}^N A_k(t)\\\nonumber
R_0&=\int_{t_{wt}}^{t_{wt}+t_{rt} } dt{1\over N}\sum_{k=1}^N R_k(t)\\\nonumber
Z_k(t)&=\left(A_k(t)-A_0\right)+\left(R_k(t)-R_0\right)i\equiv r_k(t) e^{i\phi_k(t)}\\\nonumber
\mathcal{F}&={1\over t_{rt}}\int_{t_{wt}}^{t_{wt}+t_{rt} } dt{1\over N}\sum_{k=1}^N e^{i\phi_k(t)}
\end{align}

where $t_{wt}$ and $t_{rt}$ are respectively the waiting time and run time of the simulation (see Appendix \ref{app:ForwardIntegration}). 

Figure \ref{fig:SamplesForCoherence} provides a pictorial view of how $\mathcal{F}$ corresponds to coherence among a population of cells. It is easy by eye to pick coherence for populations with $\mathcal{F}\gtrsim 0.6$, whereas smaller values seem incoherent.

Finally note that for a deterministic silent population this measure is ill-defined and will be equal to $1$. But, since in all of our multicellular simulations noise is present, we instead have $\mathcal{F}\approx 0$ whenever cells are not oscillating. 

\section{Reaction Diffusion Simulations}
\label{app:ReactionDiffusion}
The spatial simulations were done using Euler method with Neumann boundary conditions. The spatial grid spacing was $\Delta x=0.5$ and time steps were chosen according to Neumann stability criterion, $dt={\Delta x^2\over 8}$. The initial conditions were set by laying a coarser grid of different sizes on top of the simulation box and setting random values for $A$ and $R$ within each cell of the coarse grid. Initially $S$ was set to zero across space. Simulations were run for some period of time until patterns appeared.

The intersection points on the coarse grid, where a single point has four neighbors with different values, serve as the possible seeds for spiral centers. Hence a $2\times 2$ coarse grid leads to at most a single spiral on the center of simulation box (figure \ref{fig:Spatial}A) and a $20\times 20$ grid lead to many more spirals (figure \ref{fig:Spatial}B). In the latter case at most $19\times 19$ spiral centers can form. However in practice, due to random choice of initial conditions, typical length scale of spirals and topological constraints, this number tends to be much smaller.

\section{Single Cell Mechanisms of Frequency Increase}
\label{app:FrequencyIncrease}
\begin{figure}[h!]
\centering
\includegraphics[width=\linewidth]{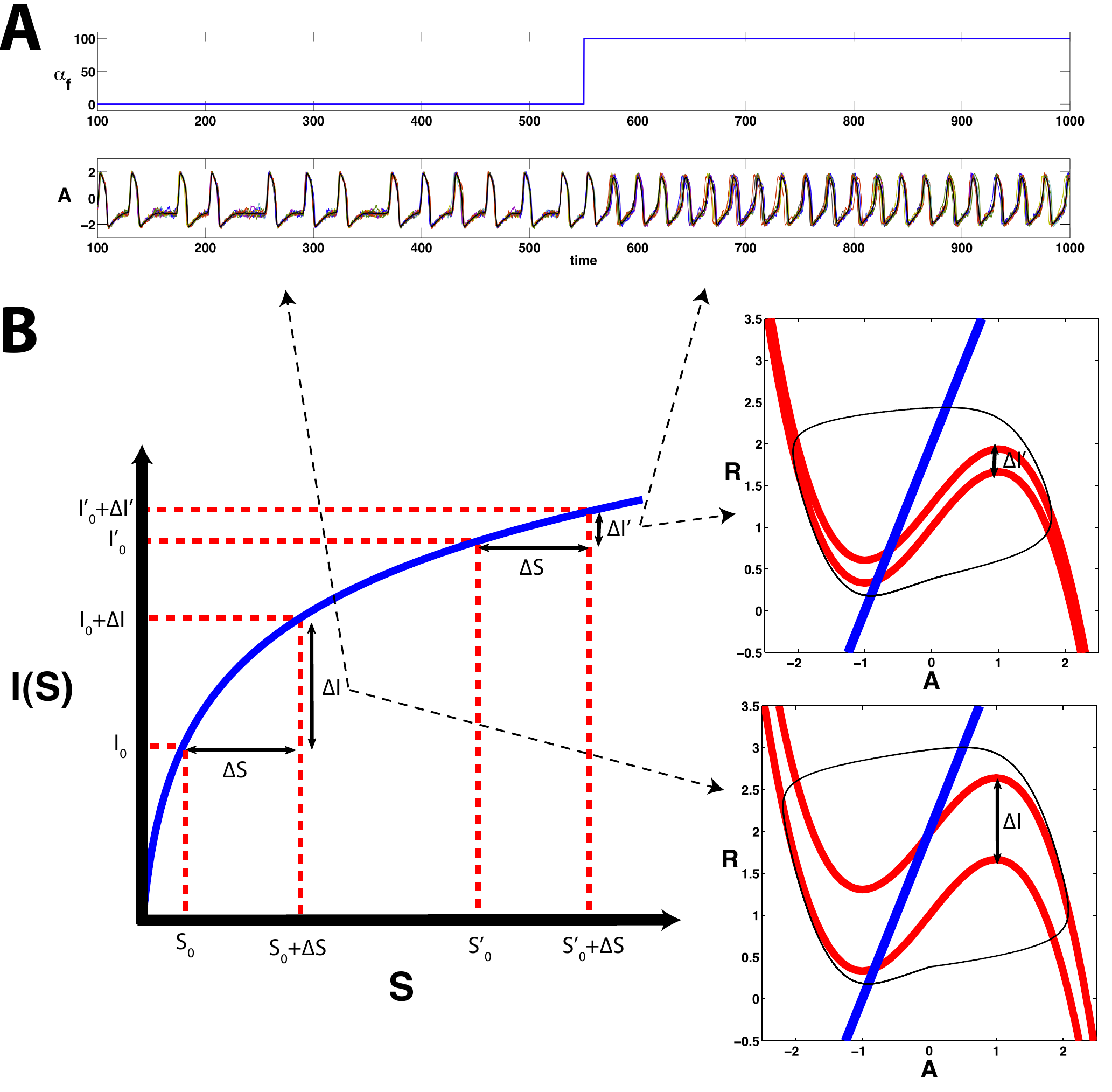}
\caption{\textbf{Frequency Increase - } {\bf A)} Time-course of a cell population in response to a step in input flow ($\alpha_f$). Parameters same as in figure \ref{fig:PhaseDiagrams}C with $J=10, \rho=1, \alpha_0=0.01, D=9000$. Midway through the simulation, $\alpha_f$ is changed abruptly from $0$ to $100$.
 {\bf B)} Blue curve shows the preprocessing function, $I(S)$, for different values of external cAMP, $S$. For two different input values, $S_0$ and $S_0'$ a constant change, $\Delta S$, leads to different changes in $I(S)$ (shown by $\Delta I$ and $\Delta I'$) such that for $S_0'>S_0$ we get $\Delta I'<\Delta I$. Phase portraits corresponding to $\Delta I$ and $\Delta I'$ are shown on the right side, showing a smaller distance between the two nullclines in the latter case and a consequent shorter trajectory over a period of oscillation. The trajectory of the system alternates between two cubic nullclines (red curves) leading to an effectively longer trajectory for larger $\Delta I$.}
\label{fig:FrequencyIncrease}
\end{figure}

\begin{figure}
\centering
\includegraphics[width=0.6\linewidth]{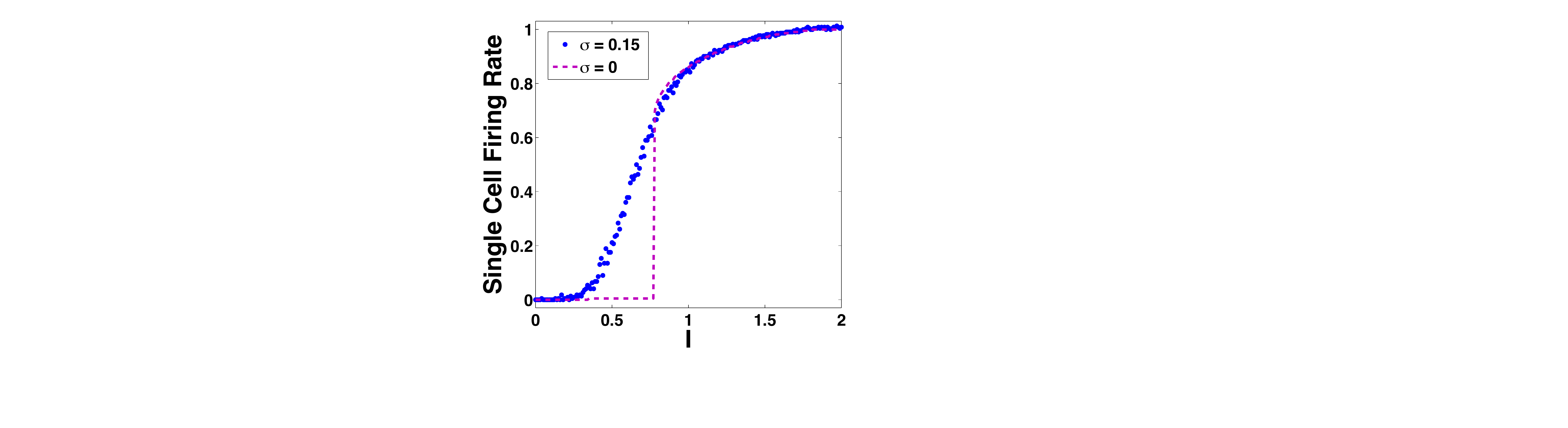}
\caption{\textbf{SI - Single Cell Firing Rate - } Firing rate as a function of constant input $I$ for noisy ($\sigma=0.15$) and noiseless ($\sigma=0$) FHN. Parameters same as in figure \ref{fig:PhaseDiagrams}C.}
\label{fig:SingleCellFreq}
\end{figure}

The mechanism introduced in section \ref{sec:FrequencyIncrease} is not the only reason for the frequency increase observed in figure \ref{fig:FrequencyIncrease}A. There is in fact a single cell frequency change that should not be confused with what has been described here. This single cell effect can be further separated into a deterministic component and a stochastic component. Figure \ref{fig:SingleCellFreq} shows the response of a single FHN to an input $I$ with and without noise. Due to the choice of nullclines for our model, an increase in $I(S)$ increases the frequency of the noiseless FHN, once the input crosses the Hopf bifurcation. Furthermore addition of noise smears the bifurcation point and creates a graded increase in frequency of single cells. As a result any flow of cAMP into a population of cells leads to a frequency increase on a single cell level that becomes amplified by the multicellular mechanism described above.
\bibliography{References.bib}
\bibliographystyle{apsrev4-1.bst}

\onecolumngrid

\end{document}